      [1999/12/01 v1.4c Il Nuovo Cimento]
\documentclass[varenna,final,pdftex]{cimento}

\usepackage{revsymb}
\usepackage{bm}
\usepackage[pdftex]{graphicx}
\usepackage{amsfonts}
\usepackage[numbers,sort]{natbib}
\usepackage{bibentry}

\def\abs#1{\left| #1\right|}

\def\gtap{\mathop{\raisebox{-.8ex}{\rlap{$\sim$}} 
\raisebox{.4ex}{$>$}}}

\newcommand{\gev}{\ensuremath{\hbox{ GeV}}}
\newcommand{\tev}{\ensuremath{\hbox{ TeV}}}

\newcommand{\fb}{\ensuremath{\hbox{ fb}}}

\title{Looking into Particle Production\\ at  the Large Hadron Collider}
\author{Chris Quigg\thanks{Electronic address: quigg@fnal.gov.}}
\institute{Fermi National Accelerator Laboratory, Batavia, Illinois 60510 USA}
\PACSes{\PACSit{13.85.Hd}{\hfill \textsf{FERMILAB--CONF--10/055--T}}
}
\begin{document}
\bibliographystyle{acm}
\maketitle

\begin{abstract}
Lightly triggered events may yield surprises about the nature of ``soft'' particle production at LHC energies. I suggest that event displays in coordinates matched to the dynamics of particle production (rapidity and transverse momentum) may help sharpen intuition, identify interesting classes of events, and test expectations about the underlying event that accompanies hard-scattering phenomena.
\end{abstract}

\section{Introduction}
\subsection{Early Running at the LHC}
At this 2010 La Thuile meeting, we have heard accounts of the first analyses of proton-proton collisions in CERN's Large Hadron Collider, at energies of $450\gev$ and $1.18\tev$ per beam.\footnote{See the talks by Fabiola Gianotti, Andrey Golutvin, Paolo Meridiani, Francesco Prino, and Andreas Wildauer, and the first publications from ALICE~\cite{Collaboration:2009dt}, ATLAS~\cite{Collaboration:2010rd}, and CMS~\cite{Collaboration:2010xs}.}
On 30 March 2010, the LHC experiments observed first collisions at $3.5\tev$ per beam, commencing a program that aims to deliver $1\fb^{-1}$ of integrated luminosity by the end of 2011. With each step in energy beyond the Tevatron Collider's $\sqrt{s} = 1.96\tev$, the LHC experiments will open new worlds.

During early low-luminosity running, the experiments will record significant numbers of lightly triggered events. Later in the run, more selective triggers will dominate the data taking. What is true of the search for the agent of electroweak symmetry breaking and other new phenomena to be sought in hard-scattering events is also true for the minimum-bias events that will dominate the early samples:\\[6pt] \centerline{\textit{We do not know what the new wave of exploration will reveal.} }

\vspace*{6pt}
The staged commissioning of the Large Hadron Collider offers the chance to map the gross features of particle production over a wide energy range. I would identify three goals: (i)~ Validate assumptions that underlie searches for new phenomena in hard-scattering events. (ii)~Develop intuition for LHC experimenters (many of whom had never---or not since the S$\bar{p}p$S collider experiments---seen two protons hit until 23 November 2009) and for interested theorists. (iii)~Make the most of the opportunity for exploration and discovery. This talk supplements my recent note~\cite{Quigg:2010nn} on this subject, where additional references may be found.

\subsection{Ken Wilson's Ancient Program} To orient ourselves, it is useful to look back to the early studies of multiple production in the 1970s. Exploration of the terrain opened up by the Fermilab bubble chambers and the CERN Intersecting Storage Rings was catalyzed, in part, by Ken Wilson's celebrated paper, ``Some Experiments in Multiple Production''~\cite{wilson}. Wilson's ``experiments'' amounted to a catalogue of informative plots to address incisive questions.

\begin{enumerate}
\setlength{\itemsep}{1pt}
\item \textit{Topological cross sections:} Do multiplicity distributions exhibit a two-component structure, suggestive of diffractive plus multiperipheral production mechanisms?
\item \textit{Feynman scaling:} Is the single-particle density $\rho_1(k_z/E, k_\perp, E)$ independent of the beam energy $E$, when plotted in terms of Feynman's scaling variable $x_{\mathrm{F}} \equiv k_z/E$?
\item \textit{Factorization:} Is the single-particle density $\rho_1(k_z/E, k_\perp, E)$  in the backward (proton) hemisphere independent of the projectile (the same for $\pi p$ and $pp$ scattering)?
\item \textit{$dx/x$ spectrum:} Does the single-particle density exhibit a flat plateau in the central region when plotted in terms of the rapidity, $y \equiv \frac{1}{2}\ln{[(k_0 + k_z)/(k_0 - k_z)]}$ ?
\item \textit{Correlation length experiment:} Does the two-particle correlation function $C(y_1,y_2) \equiv \rho_2(y_1,y_2)-\rho_1(y_1)\rho_1(y_2)$ display short-range order, $\propto \exp(-\abs{y_1 - y_2}/L)$?
\item \textit{Factorization test (\#3) with central trigger} (to eliminate diffraction).
\item \textit{Double Pomeron exchange:} Do some events display low central multiplicity with large rapidity gaps on both ends?
\end{enumerate}
The experimental studies responded affirmatively to  questions 1--6. The CDF Collaboration has recently reported the production of isolated charmonium states in the central region, characteristic of the reaction $\mathbb{PP} \to \chi_{c0}$~\cite{Aaltonen:2009kg}, as anticipated in question 7.

\section{Particle Production at the LHC}
\subsection{New phenomena ahead?}
This doesn't mean, however, that ``soft'' particle production should be regarded as settled knowledge. It has not yet been exhaustively studied at the Tevatron (see~\cite{Aaltonen:2009ne,Aaltonen:2010rm} for recent important progress), and so we can't be sure that what was inferred from experiments up to $\sqrt{s} = 63\gev$ accounts for all the important features at Tevatron energies. At the highest energies, well into the ($\propto \ln^2{s}$?) growth of the $pp$ total cross section, long-range correlations might show themselves in new ways. {The high density of partons carrying  $p_z = 5\hbox{ to }10\gev$ may give rise to hot spots in the spacetime evolution of the collision aftermath, and thus to thermalization or other phenomena not easy to anticipate from the QCD Lagrangian.} We might anticipate a growing rate of multiple-parton interactions~\cite{Bernhard:2010su}, perhaps involving correlations among partons. For example, the quark-diquark component of the proton might manifest itself in elementary collisions involving diquarks. The $\ln{s}$ expansion of the rapidity plateau softens kinematical constraints in the central region, and the sensitivity to high-multiplicity events  (or otherwise rare occurrences) of modern experiments vastly exceeds what could be seen with bubble-chamber statistics. 
The CMS Collaboration reports~\cite{Collaboration:2010xs} that the standard \textsc{pythia} tunes underestimate the growth with energy of the central density of charged particles, $dN_{\mathrm{ch}}/d\eta|_{\eta = 0}$, from $\sqrt{s} = 900\gev$ to $2.36\tev$. At $\sqrt{s} = 900\gev$, the ATLAS experiment observes~\cite{Collaboration:2010rd} that  $dN_{\mathrm{ch}}/d\eta|_{\eta = 0}$ lies some (5 - 15)\% above the predictions of the Monte Carlo models.  For all these reasons, I suspect that a few percent of minimum-bias events collected at $\sqrt{s} \gtap 1\tev$ might display unusual event structures.\footnote{Many of the questions posed in the FELIX physics document~\cite{Ageev:2001qv} are apt for the detectors now taking data at the LHC.} We should look! But how?

\subsection{Learning to See}
\begin{figure}[tb]
\centerline{\includegraphics[width=0.7\textwidth]{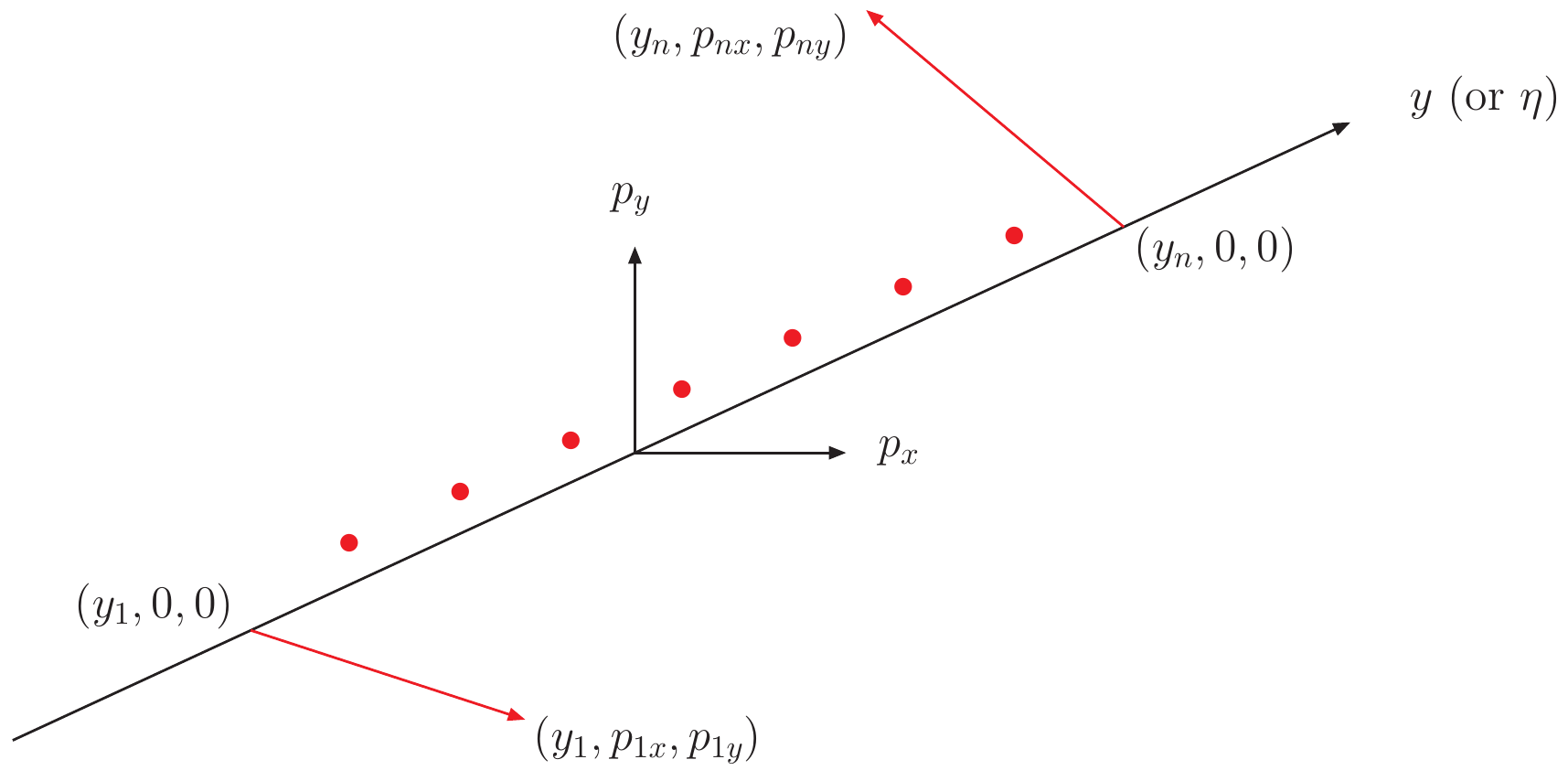}}
\caption{Schematic event display in $(y,\vec{p}_\perp)$ space. \label{fig:example}}
\end{figure}

I believe that \textit{looking at events} can be an important part of the answer. Blind analysis~\cite{Klein:2005di} has won a secure place in our practice of particle physics, as a talisman against experimenter's bias, but it is not apposite when we are seeking to get the lay of the land. It would be a big mistake to suppose that we know all the important questions, even before we arrive in the new world!
Bjorken suggested long ago~\cite{Bjorken:1971ww} a three-dimensional representations of multiparticle events that could engage our human powers of visualization and pattern recognition, in the hope of identifying important new questions. For particle production in soft collisions, it is not spatial coordinates that are most apt, but a representation in terms of (pseudo)rapidity and (two-dimensional) transverse momentum. To begin, draw a (pseudo)rapidity axis as an oblique line. Represent each track $i$ in the event by a vector drawn from $(y_i,0,0)$ to $(y_i,p_{ix},p_{iy})$, as in the example shown in Figure~\ref{fig:example} (all scales linear).
The $(y,p_\perp)$ representation is none other than a curled-up vector representation of the LEGO$^{\scriptsize \textregistered}$ plot for individual tracks, with thresholds for display set as low as possible. 
\begin{quote}
As a start, I encourage the LHC collaborations to produce $(y,\vec{p}_\perp)$ displays of minimum-bias events acquired during early running. Samples as small as a few hundred events would already build intuition, but I would go further. I suggest that the collaborations make available live streams of $(y,\vec{p}_\perp)$ representations, along with the online displays of events that show the structure in terms of detector elements in ordinary space.  It is useful to color the tracks to label their charges, and to identify species where possible.\\[6pt]\textit{More is to be learned from the river of events than from a few specimens!} \\[6pt]Changes in event structure as a function of beam energy, or the onset of new features, might raise important questions.
\end{quote}

\begin{figure}
\centerline{\includegraphics[width=\textwidth]{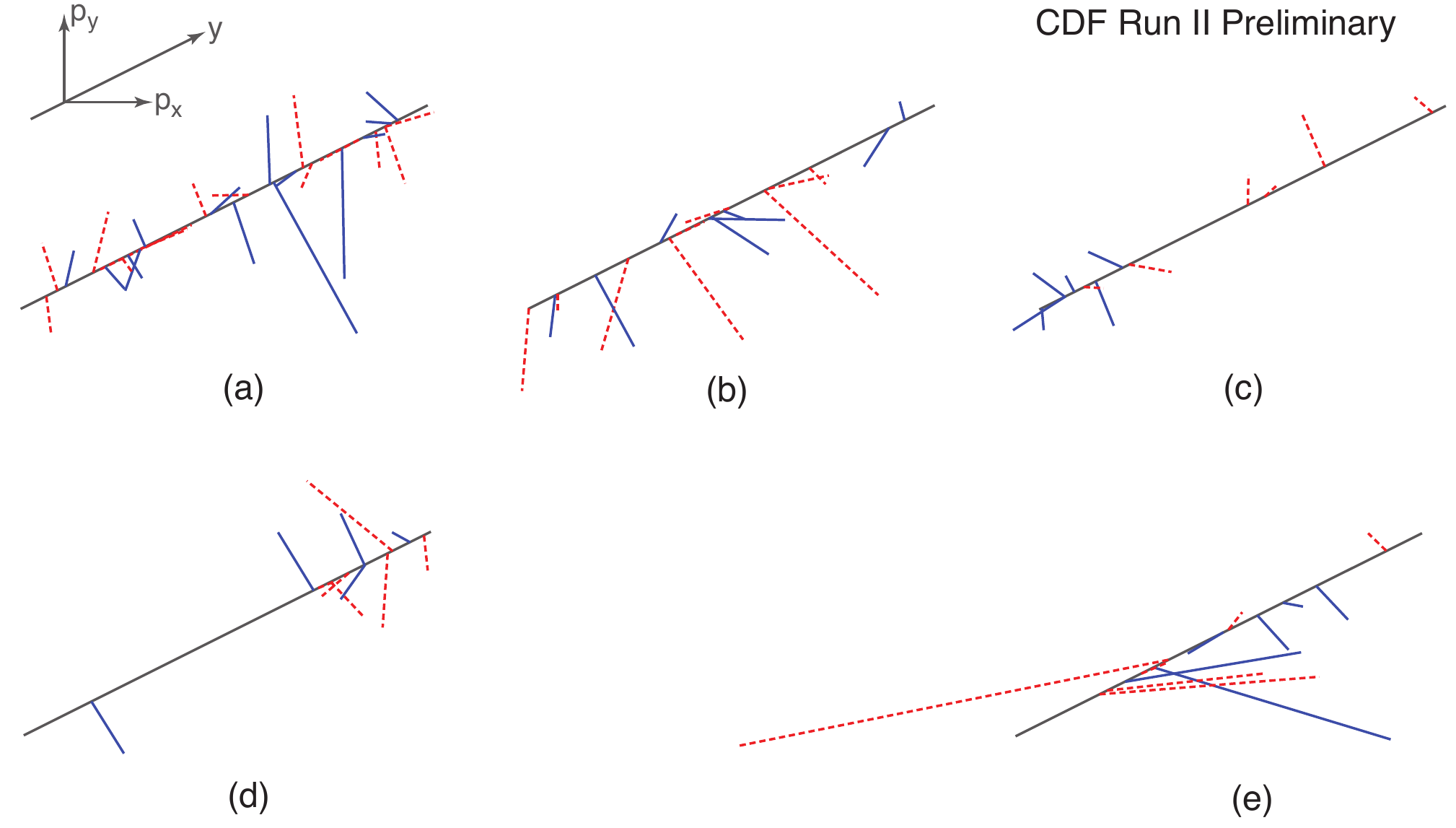}}
\caption{Example $(y,\vec{p}_\perp)$ event displays from Run~2 of the CDF Experiment at Fermilab. These are chosen from a sample of ``zero-bias'' events with at least 10 tracks in the central region and one good primary vertex. Each rapidity axis spans $-1 \le y \le 1$; the length of the transverse-momentum axes is $1\gev$. Dashed red and solid blue lines label positively and negatively charged tracks, respectively. (a) Local compensation of $\vec{p}_\perp$ and electric charge. (b) Local $\vec{p}_\perp$ imbalance. (c) Local charge imbalance. (d) Rapidity gap. (e) Hot spot.
\label{fig:cdfevents}}
\end{figure}

Thanks to work of Niccol\`{o} Moggi~\cite{Moggi:2009lt} and William Wester, I can show in Figure~\ref{fig:cdfevents} a few example $(y,\vec{p}_\perp)$ displays of events recorded by the CDF Collaboration in Run~2 at the Tevatron, in $\bar{p}p$ collisions at $\sqrt{s} = 1.96\tev$. The events shown there are chosen from a ``zero-bias'' sample after selections to ensure a single primary vertex within 30~cm ($1\sigma$) of the nominal crossing point and require (for visual interest) at least 10 well-matched tracks in the rapidity interval $-1 \le y \le 1$.
(By selecting higher-than-average-multiplicity events, we are excluding candidates for $\gamma\gamma$ and double-Pomeron events that would be restricted to low central multiplicities.)
Event (a) exhibits the average behavior familiar at lower energies, whereby transverse momentum and additive quantum numbers such as electric charge are compensated locally in rapidity. Each of the remaining events deviates from the typical expectations, inviting further study of large numbers of events to ascertain whether they fit neatly into fluctuations about the mean or suggest new event classes. Transverse momentum is unbalanced in event (b); positive and negative charges are separated in event (c); a rapidity gap of slightly more than one unit appears in event (d); and a good deal of the action in event (e) seems concentrated in a ``hot spot'' in rapidity.

By scanning many events, it should be possible to (quickly) develop intuition about what is ``normal,'' both for lightly triggered events and for events that satisfy a hard trigger. It will also be valuable to compare streams of real events with streams of simulated events. It is certain that something is missing from the Monte Carlo programs. We need to learn what the omissions are, and how important they are to our understanding. Attentive scanning could well yield the suggestion of unanticipated phenomena---at the level of a few percent, one in a thousand, even one in $10\,000$. Modern computer tools make it straightforward to construct $(y,p_\perp)$ displays that can be zoomed, panned, and rotated in three dimensions. The ability to manipulate events and regard them from changing perspectives can engage our human powers of perception more fully.

\subsection{New Physics in the Weeds}
The strong interactions are extraordinarily rich. Even as we learn to extend the reach of perturbative QCD beyond reactions involving a few partons in the final state, we should be attentive to the whole range of strong-interaction phenomena. The rest of the story includes common processes with large cross sections such as ``soft'' particle production, elastic scattering, and diffraction. It may well be that interesting, \textit{unusual} occurrences happen outside the framework of perturbative QCD---happen in some collective, or intrinsically nonperturbative, way. A powerful technique to isolate hard-scattering reactions is to impose stringent cuts in the data selection, or to clarify the essential structure of events by setting display thresholds high. When scanning event displays for hints of new phenomena, however, it may be advantageous to set the display thresholds \textit{as low as possible.}

\begin{figure}[tb]
\centerline{\includegraphics[width=0.7\textwidth]{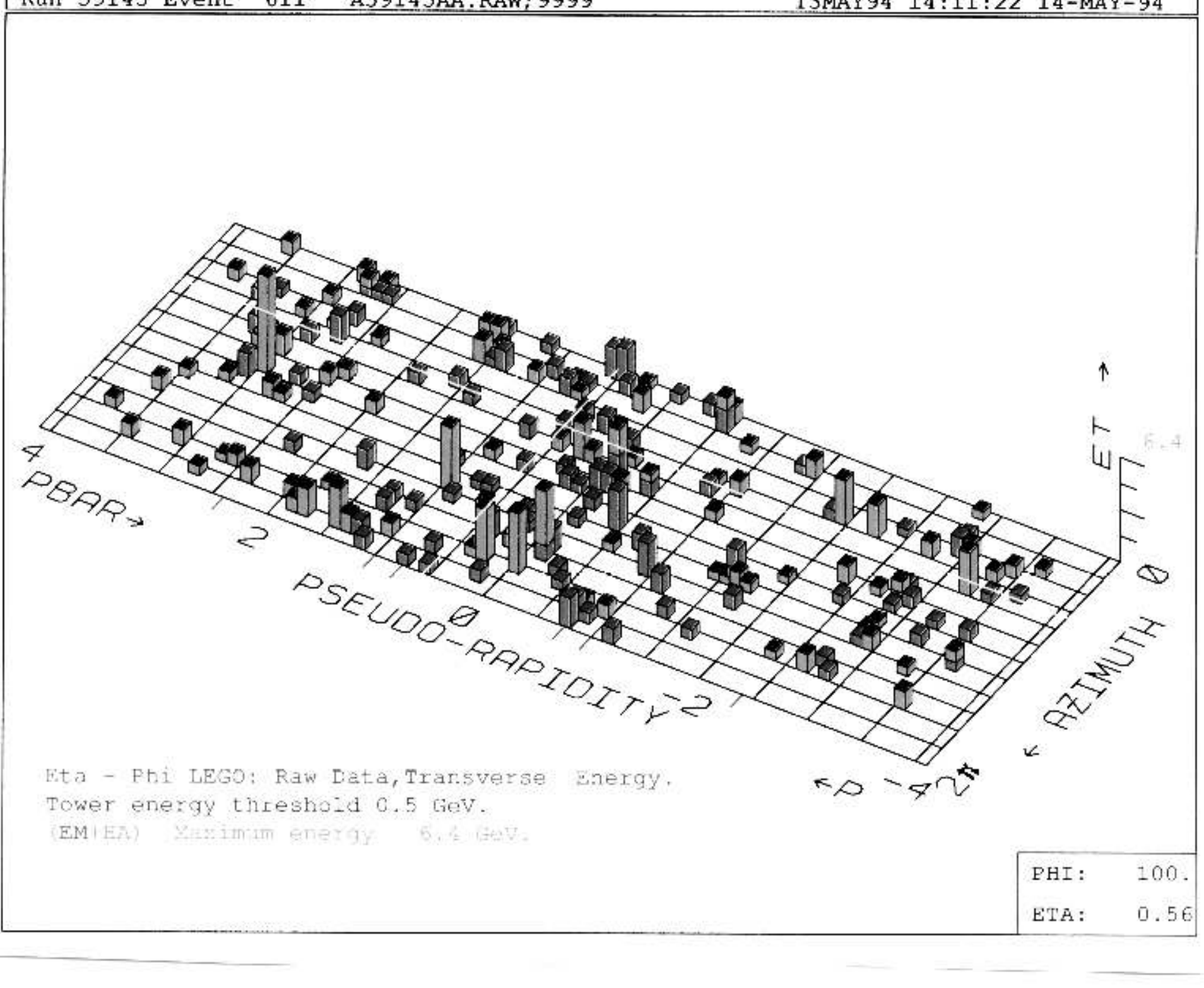}}
\caption{An unusual event captured in the CDF Run 1 detector.  \label{fig:drasko}}
\end{figure}

An interesting example---an \textit{atypical} event observed in $\bar{p}p$ interactions at $\sqrt{s} = 1.8\tev$ by CDF's Run 1 detector,  is shown in Figure~\ref{fig:drasko}.\footnote{See Figure~3(c) of~\cite{Albajar:1987sa} for a similarly isotropic event recorded in the UA1 Detector in $\bar{p}p$ collisions at $\sqrt{s} = 630\gev$, in which $\sum E_\perp = 209\gev$ for $\abs{y} < 1.5$.} This event was accepted by a $\sum E_\perp$ trigger, without any topological requirement. The LEGO$^{\scriptsize \textregistered}$ plot shows many bursts of energy:
More than a hundred active towers pass the display threshold of $0.5\gev$. The total transverse energy
in the event is $321\gev$, but it is not concentrated in a few sprays, it is everywhere. The central
tracking chamber records about sixty charged particles.

I am assured that this ``hedgehog'' event is authentic; it is not merely coherent noise in the counters. The colleague who selected this specimen estimated similar events to be about as common in the online event stream as $Z^0$ production and decay into
lepton pairs: about one in ten thousand triggers. I include this outlier as a reminder
that when we think about the strong interactions outside the realm of a single hard scattering,
we should think not only about the large diffractive and ``multiperipheral'' cross sections, but also about less common phenomena.

\section{Opportunities for Exploration and Discovery}

The minimum-bias and lightly triggered data recorded during early LHC running will be valuable for developing intuition and for validating the assumptions that underlie searches for new physics in hard-scattering events. However, these data sets, to be gathered over steps in beam energy, also represent an important opportunity for exploration and discovery. One promising track will be to emulate the early studies of multiple production, which emphasized observables constructed from individual particles: topological cross sections (multiplicity distributions, including forward-backward asymmetries of multiplicity distributions), inclusive and semi-inclusive two-particle correlation functions, and charge-transfer studies. Some measurements that would be especially informative for  refining Monte Carlo event generators are suggested in~\cite{SkandsMB2010}. For some classes of events, analyses of bulk properties, such as studies of elliptic flow and determinations of thermodynamic parameters may prove powerful. We will need all the established methods---plus novel techniques---to learn to see what the LHC data have to show.

It is not too late to characterize particle production more completely in the Tevatron experiments. The existing samples of lightly triggered events can be mined further, with an eye to establishing in detail the mechanisms at play in particle production and identifying suggestive classes of unusual events. It is worth considering a brief, dedicated, Tevatron run at $\sqrt{s} = 900\gev$, to match the samples collected in the LHC's pilot run at the end of 2009. The similarities and differences between $pp$ and $\bar{p}p$ collisions may be revealing.

I advocate looking at individual events, not just distributions.
Beyond honing intuition, the first effect of \textit{looking at events,} displayed in appropriate coordinates, may be to validate in broad terms the prevailing picture of particle production. We should also be able to test the completeness of the Monte Carlo frameworks that have become so indispensable to the search for new (hard-scattering) phenomena. 
I think it likely that we will encounter suggestions of new event classes, to be pursued in focused studies that go beyond visual inspection. New trends may emerge with increasing beam energy, or at the extremes of high and low multiplicity. The comparison of events with and without a hard trigger should be revealing. The goal of the visual approach  is to discover as completely as we can the richness of phenomena that our theories will have to explain, and to orient us for detailed exploration of the new worlds.

\acknowledgments
I thank Niccol\`{o} Moggi and William Wester for the example event displays from CDF Run~2, and the CDF QCD convenors, Christina Mesropian and Sasha Pranko, for permission to include them in this presentation. Drasko Jovanovic provided the hedgehog event in Figure~\ref{fig:drasko}, and Jim Rohlf called my attention to the earlier UA1 observations. I thank James Bjorken for many stimulating conversations. I am grateful to the organizers of these Rencontres de Physique de la Vall\'{e}e d'Aoste for their kind invitation to take part.
Fermilab is operated by Fermi Research Alliance, LLC  under Contract
No.~DE-AC02-07CH11359 with the United States Department of Energy.  

\bibliography{QuiggL2C}

\end{document}